# MICRO AND NANO-PATTERNING OF SINGLE-CRYSTAL DIAMOND BY SWIFT HEAVY ION IRRADIATION

G. García[1,*], I. Preda[1,+], M. Díaz-Híjar[2,3], V. Tormo-Márquez[3],

O. Peña-Rodríguez[4], J. Olivares[2,3], F. Bosia[5,6], N.M. Pugno[7,8,9], F. Picollo[5,6], L. Giuntini[10],

A. Sordini[11], P. Olivero[5,6], L. López-Mir[12], C. Ocal[12]

[1] ALBA Synchrotron Light Source (CELLS-ALBA), 08290, Cerdanyola del Vallès, Barcelona, Spain.
[2] Instituto de Óptica, Consejo Superior de Investigaciones Científicas (CSIC), C/Serrano 121, E-28006 Madrid, Spain.
[3] Centro de Microanálisis de Materiales (CMAM), Universidad Autónoma de Madrid (UAM), Cantoblanco, E-28049 Madrid, Spain.
[4] Instituto de Fusión Nuclear (UPM), C/ José Gutiérrez Abascal 2, E-28006 Madrid, Spain.
[5] Physics Department and "Nanostructured Interfaces and Surfaces" (NIS) Inter-departmental Centre, University of Torino, Torino, Italy.
[6] INFN - National Institute of Nuclear Physics, Section of Torino, Torino, Italy.
[7] Laboratory of Bio-Inspired & Graphene Nanomechanics, Department of Civil, Environmental and Mechanical Engineering, University of Trento, Trento, Italy.
[8] Centre of Materials and Microsystems, Bruno Kessler Foundation, Trento, Italy.
[9] School of Engineering and Materials Science, Queen Mary University, London, UK.
[10] Istituto Nazionale di Fisica Nucleare, Sezione di Firenze, Sesto Fiorentino, Italy.
[11] National Institute of Optics (INO-CNR), Firenze, Italy.
[12] Institut de Ciència de Materials de Barcelona (ICMAB-CSIC), Campus de la UAB, 08193 Bellaterra, Barcelona, Spain.

[*] Corresponding author. Tel: +34 935924300. E-mail: ggarcia@cells.es (Gastón García).
[+] Presently at MAX IV Laboratory, Fotongatan 2, 225 94 Lund, Sweden.

**Abstract**

This paper presents experimental data and analysis of the structural damage caused by swift-heavy ion irradiation of single-crystal diamond. The patterned buried structural damage is shown to generate, via swelling, a mirror-pattern on the sample surface, which remains largely damage-free. While extensive results are available for light ion implantations, this effect is reported here for the first time in the heavy ion regime, where a completely different range of input parameters (in terms of ion species, energy, stopping power, etc.) is available for customized irradiation. The chosen ion species are Au and Br, in the energy range 10-40 MeV. The observed patterns, as characterized by profilometry and atomic force microscopy, are reported in a series of model experiments, which show swelling patterns ranging from a few nm to above 200 nm. Moreover, a systematic phenomenological modelling is presented, in which surface swelling measurements are correlated to buried crystal damage. A comparison is made with data for light ion implantations, showing good compatibility with the proposed models. The modelling presented in this work can be useful for the design and realization of micropatterned surfaces in single crystal diamond, allowing to generate highly customized structures by combining appropriately chosen irradiation parameters and masks.

1. Introduction

Structural damage induced in single-crystal diamond by ion irradiation has been studied in a variety of experimental configurations, which mostly include the use of medium/light ions at ~0.1–1 MeV energies for both fundamental studies [1-5] and device applications [6-9]. Remarkably, no systematic irradiation studies with swift heavy ion beams have been performed until very recently [10]. For all the data presented in this work, the damage generation mechanism can be attributed exclusively to nuclear stopping, since the electronic stopping force lies in the range below 14 keV/nm [2,3,10]. Due to the energy dependence of nuclear stopping, ion beams with high enough energy generate significant structural damage below the sample surface, whereas the surface layers undergo limited structural modifications. The length scales involved are typically in the micrometer range, both for the thickness of the undamaged surface layer and for that of the buried damaged one. The effect of the induced stress on the crystalline surface layer, generated by the expansion of the underlying damaged volume, gives rise to surface swelling, which has been observed and phenomenologically described in the light ion regime or at low ion energies [11-13].

The aim of this paper is twofold: firstly, to report the swelling effect in the swift heavy ion regime, comparing experimental results with the phenomenological model developed for light ions [12, 13] in order to assess its validity also for swift heavy ions and, secondly, to highlight the potential exploitation of the swelling effect, with an extended range of input parameters offered by swift heavy ions of arbitrary species, to generate customized surface landscapes of lightly damaged diamond crystals with interesting aspect ratio characteristics. Phenomenological models are described in order to provide simple tools to fine-tune irradiation parameters to the desired surface effect in a customized way.

## 2. Swift heavy ion implantations

Optical-grade single-crystal diamond samples, (3×3×0.3) mm$^3$ in size, (100) oriented and with two polished surfaces were supplied by ElementSix [14]. The samples were classified as type IIa, corresponding to concentrations of N and B impurities below 1 ppm and 50 ppb, respectively. Irradiations were performed at CMAM [15, 16], using the standard beamline [17]. Samples were implanted in frontal geometry on their polished surfaces, with slight tilting in order to avoid channelling effects. The different ion beams employed were defocused so as to provide homogeneous irradiation of the whole sample surface. Homogeneity was carefully tested for each beam configuration by irradiating a test quartz sample and measuring the induced luminescence on a CCD camera.

The adopted beams included Au and Br ions, with energies in the range 10-40 MeV and fluences in the range from $5\times10^{13}$ cm$^{-2}$ to $5\times10^{14}$ cm$^{-2}$. During the corresponding irradiations, samples were masked with suitable grids, providing lateral 2D irradiation patterns to test the material response. Beam currents from a few tens of nA to 130 nA where used as available from the CMAM accelerator for the different ion species and energies chosen. The different experimental configurations are summarized in Table 1. The chosen ion species and energies are focused on a systematic study as a function of fluence and with different mask configurations for 10 MeV Au ions, complemented with a sample at higher Au beam energy and with a few samples irradiated with Br, exploring a lower nuclear stopping power range. Br irradiation parameters were chosen so as to provide some overlap with the information provided by Au irradiations, as discussed in section 4 below.

| Ion species | Ion energy [MeV] | Ion fluence [$10^{13}$ cm$^{-2}$] | Mask mesh / irradiated square side |
|---|---|---|---|
| Au | 18.6 | 10 | 1 mm / 0.97 mm |
| Au | 10.0 | 5 | 1 mm / 0.97 mm |
| Au | 10.0 | 10 | 1 mm / 0.97 mm |
| Au | 10.0 | 25 | 1 mm / 0.97 mm |
| Au | 10.0 | 50 | 1 mm / 0.97 mm |
| Au | 10.0 | 5 | 80 µm / 60 µm |
| Au | 10.0 | 50 | 80 µm / 60 µm |
| Br | 36.7 | 20 | 80 µm / 60 µm |
| Br | 40.0 | 5 | 11 µm / 5.5 µm |
| Br | 40.0 | 10 | 11 µm / 5.5 µm |
| Br | 40.0 | 45 | 11 µm / 5.5 µm |

Table 1: *Experimental configurations for swift heavy ion irradiation of single-crystal diamond.*

The sample surface topography of as-irradiated samples was characterized by profilometry, used in this paper as the main characterization technique for systematic analysis of the swelling effect. The measurements were conducted at the Nanoquim Platform Laboratory at Institut de Ciència de Materials de Barcelona (ICMAB-CSIC) using a Profilometer P16+ from KLA Tencor. In order to validate the results and provide further insight into the sample morphologies, atomic force microscopy (AFM) measurements were also carried out on two selected samples at ICMAB-CSIC with a MFP3D Asylum equipment, using Silicon tips (radius 9 ± 2 nm) mounted on levers with nominal stiffness constant k = 2 N m$^{-1}$ (AC240TS). Experimental details and measuring set-up can be found elsewhere [18] and AFM data were analyzed using the WSxM free software [19].

Irradiated areas of the diamond samples were found to appear opaque after ion irradiation for all cases given in Table 1. Therefore, an optical micrograph clearly shows the irradiation pattern as generated by the mask used in each case, as shown in Fig. 1a. For this implantation, the selected mask allowed to irradiate 5.5×5.5 µm$^2$ squares, separated by 5.5 µm from each other in both perpendicular directions.

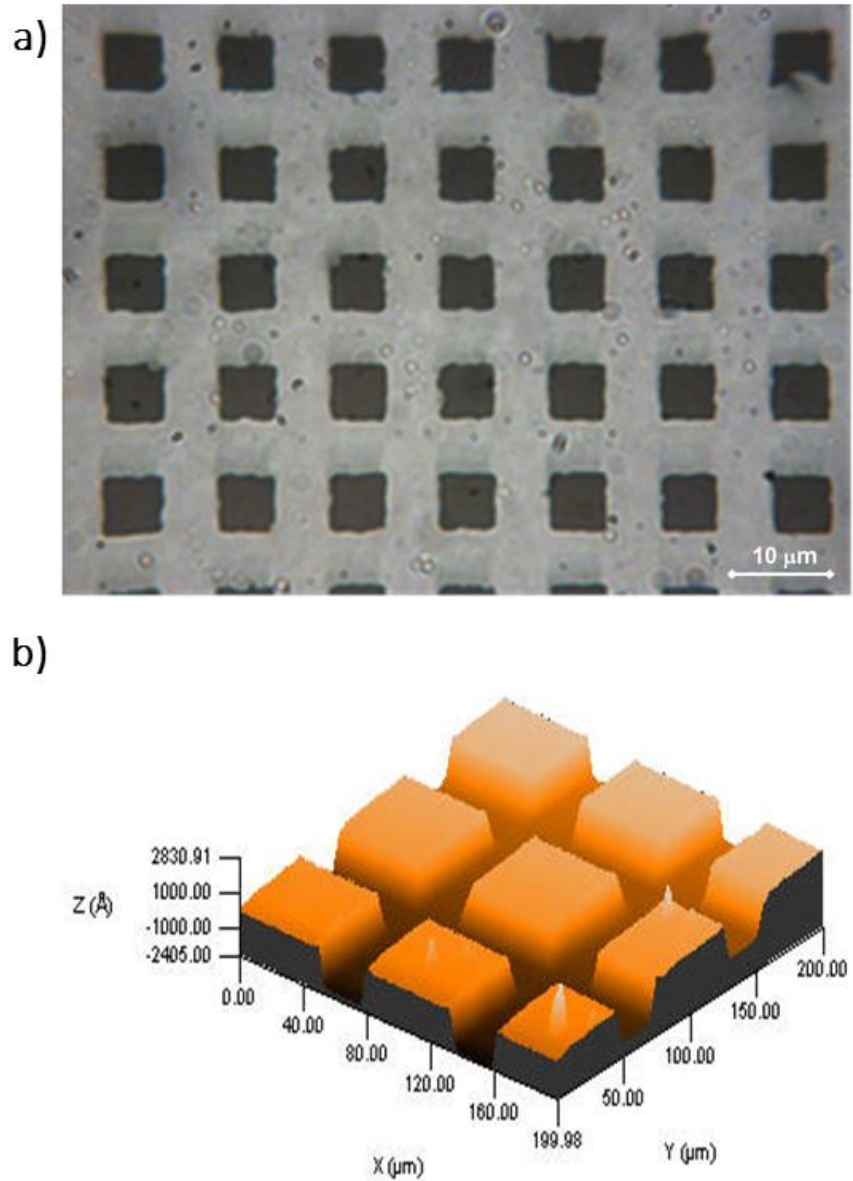

Fig. 1: a) *Optical microscope image of a diamond sample after irradiation with a 40 MeV Br beam at a fluence of $5\times10^{13}$ cm$^{-2}$. An 11 μm mesh mask with 5.5 μm square apertures was used for irradiation. Dark areas of 5.5×5.5 μm$^2$ correspond to the irradiated sample surface, whereas the surrounding grey area corresponds to the unirradiated surface (i.e. the mask-covered regions). b) Surface topography, as measured with a profilometer, of a single-crystal diamond sample after irradiation with a 10 MeV Au beam at a fluence of $5\times10^{14}$ cm$^{-2}$. An 80 μm mesh mask with 60 μm square apertures was used for the irradiation.*

The resulting pattern was studied in detail by means of 2D profilometer scans on a selected set of samples. Irradiated (opaque) regions have developed structural damage [10], generating buried volumes where the diamond lattice has been amorphized, leading to a density decrease and therefore a stress exerted on the surrounding crystalline diamond regions. The thin diamond slab on top of each of these modified volumes deforms and generates a swelling pattern on the surface, whereas unirradiated (transparent) areas do not deform. Irradiation using mesh masks gives rise to "landscapes" consisting of "plateaux" and "canyons", in which the latter have widths corresponding to the mask wire diameter, and a depth ranging from a few to hundreds of nanometers, depending on irradiation parameters. Fig. 1b shows an example, in which $60\times60\ \mu m^2$ plateaux, ca. 200 nm in height, are surrounded by 20 $\mu$m wide canyons.

In order to crosscheck the height and sharpness of the plateau-to-canyon steps, as complementary to the profilometry measurements, two selected samples were measured by AFM in contact mode under controlled low load ($\leq$150 nN). Since in the employed AFM set-up the accessible scanned areas are relatively small ($\leq 60\times60\ \mu m^2$), a complete plateau could not be imaged. Therefore, the monitored areas were chosen to contain at least part of two neighbouring irradiated regions plus the intermediate unirradiated gap (Fig. 2a). With respect to standard profilometry, AFM is well known to give more accurate measurements of the step abruptness, with a negligible tip-convolution effect. Fig. 2b shows the 1D profile along the segment depicted in the 2D image (Fig. 2a). The measured height is $226 \pm 5$ nm, as can be observed in the magnified step edge profile (Fig. 2c). The sharpness of the generated features is $\sim 2\ \mu$m. Additional information is obtained by measuring the root mean square (rms) surface roughness on both the surfaces corresponding to the top of the plateau and the intermediate canyon, resulting in rms estimates of $4.5 \pm 0.5$ nm and $1.8 \pm 0.2$ nm, respectively. The small increase of the rms roughness after sample irradiation by heavy energetic ions indicates that this has a limited impact on the surface roughness of the sample. A second single-crystal diamond sample was analysed by AFM after irradiation with a 36.7 MeV Br beam at a fluence of $2\times 10^{14}$ cm$^{-2}$, masked by an 80 $\mu$m mesh grid of 20 $\mu$m thick wires. In this case, the plateaux height was of $23 \pm 6$ nm, as obtained from height histograms over areas of $30\times 30\ \mu m^2$ and a step width of $\sim 5\ \mu$m.

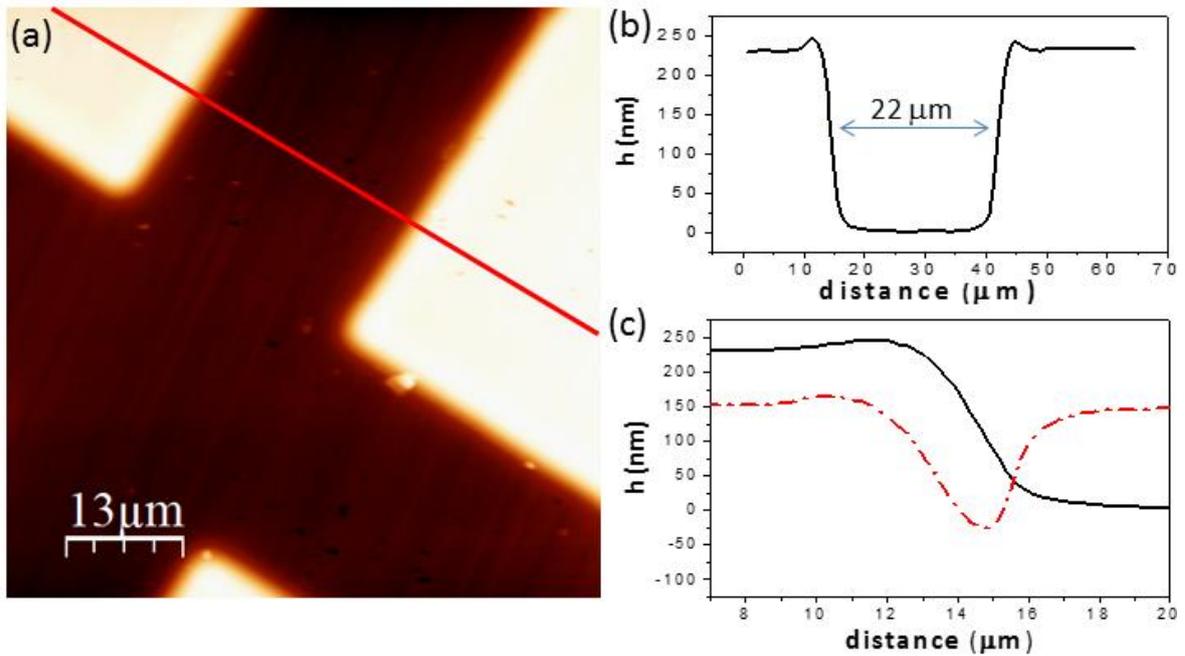

Fig. 2: a) *Topography as measured by AFM of a diamond single-crystal sample after irradiation with a 10 MeV Au beam at a fluence of $5\times10^{14}$ cm$^{-2}$. An 80 μm-mesh mask with 60 μm square apertures was used during irradiation. b) Profile along the segment indicated in the image in (a). c) Magnification of the step edge profile (black line) and its derivative (red dot-dashed line) to better visualize the abruptness of the step.*

The average height of the structures for each sample can be measured either by taking several line scans from the 2D maps or by performing height histograms as explained in the caption of Fig. 3. Thus, it is possible to correlate these values with the input irradiation parameters. Fig. 3a shows several line scans as taken from a single profilometry 2D map. The step height is seen to show uncertainties in the 10% range. Within this level of accuracy, the data are very reproducible. The result of the analysis of the height histograms is shown in Fig. 3b for AFM measurements. In this case the uncertainty of the step height measurement is lower than 3%.

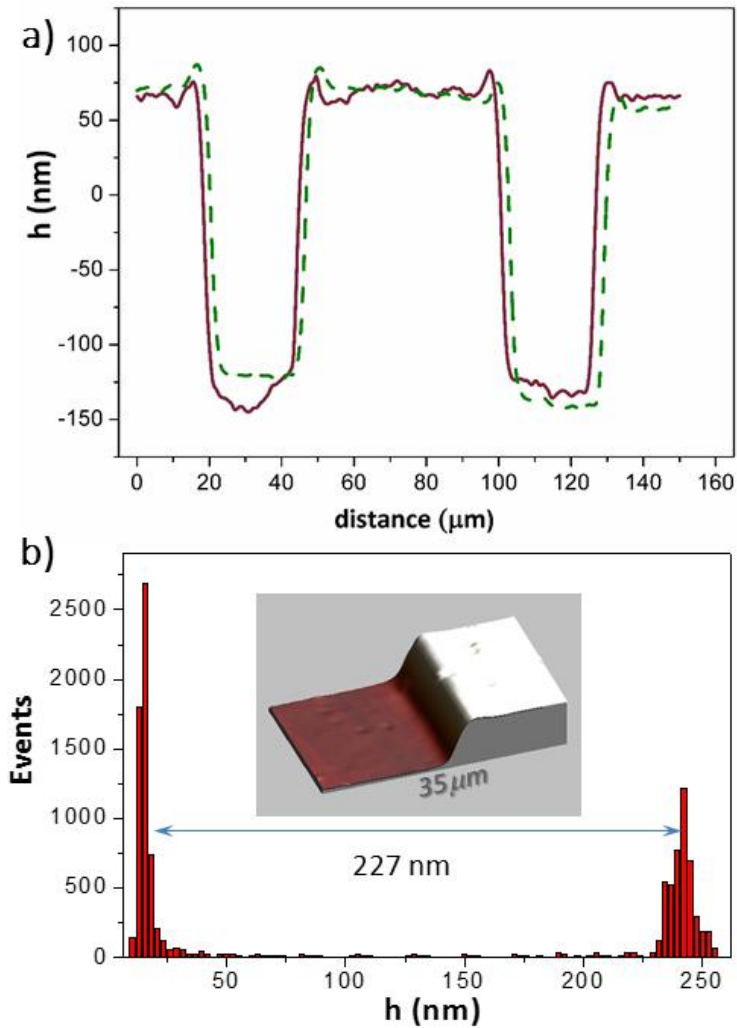

Fig. 3: a) *Line scans taken at two representative points of a 2D profilometry scan, yielding an average height $\bar{h}$ =204 nm.* b) *Height analysis. A 3D image corresponding to a selected area containing part of one structure and part of the canyon is selected to illustrate the procedure. The histogram represents the frequency (events) of each height value on the image. It shows two main peaks. The peak at the left corresponds to the base of the canyon and that at the right to the plateau. The difference in height or step height is therefore the distance between the two peak centers, i.e. $\bar{h}$ = 227 nm in this case. Both graphs correspond to the same sample. Irradiation parameters are those given in Fig. 2.*

On the other hand, the average width of the step was evaluated to be in the 2-5 μm range, both in profilometry and AFM measurements (see Figure 2c). In order to discard any effect due to scan speed, AFM measurements at different speed rates were performed. On the other hand tip-surface convolution leading to slope broadening was also ruled out by checking the scan of a commercial Si grid. The measured width is therefore attributed to the actual smooth profile of the step itself. The analysis of these data is discussed below.

### 3. Light ion implantations

Ion implantation was performed on type IIa CVD samples by ElementSix consisting in (100) oriented single crystals of size $3\times3\times1.5$ mm$^3$, with two optically polished large opposite surfaces. The samples were implanted in a broad range of fluences with different ion species and energies: $3\text{-}50\times10^{16}$ cm$^{-2}$ (He 1.3 MeV), $1\text{-}15\times10^{16}$ cm$^{-2}$ (H 2 MeV) and $0.9\text{-}8\times10^{16}$ cm$^{-2}$ (H 3 MeV). He ions were implanted at the ion microbeam line of the INFN Legnaro National Laboratories (Padova), H ion implantations were performed at the external microbeam line of the LABEC INFN facility (Firenze), and C ion implantations were performed at Ruđer Bošković Institute in (Zagreb, Croatia). In all cases, the samples were implanted in frontal geometry on their polished surfaces, with slight specimen tilting in order to avoid channelling effects. Square areas of $125\times125$ μm$^2$ were implanted by raster scanning a 20–30 μm ion beam. The implantations were performed at room temperature, with ion currents of ~1 nA. As previously discussed, the implanted areas can be visualized by means of an optical microscope, as regions of increasing opacity with increasing fluence, as shown in Fig. 4a for a sample implanted with 2 MeV H ions.

Surface swelling data were acquired at the the Istituto Nazionale di Ottica (INO) with a Zygo NewView 6000 system, which exploits white light interferometry to provide detailed measurements of 3D profiles. A vertical resolution of 0.1 nm was achieved over a lateral range up to 150 μm, while lateral resolution varied from 4.6 μm up to 0.6 μm, depending on the objective. A typical measurement is shown in Fig. 4 b), where the swelling profile of a He 1.3 MeV implanted area (for a fluence of $7.69\times10^{16}$ cm$^{-2}$) is reconstructed. Some of these data have partially been reported in the past in previous publications [12, 22], but are included here to present as complete an overview as possible of results for different ion types and energy ranges.

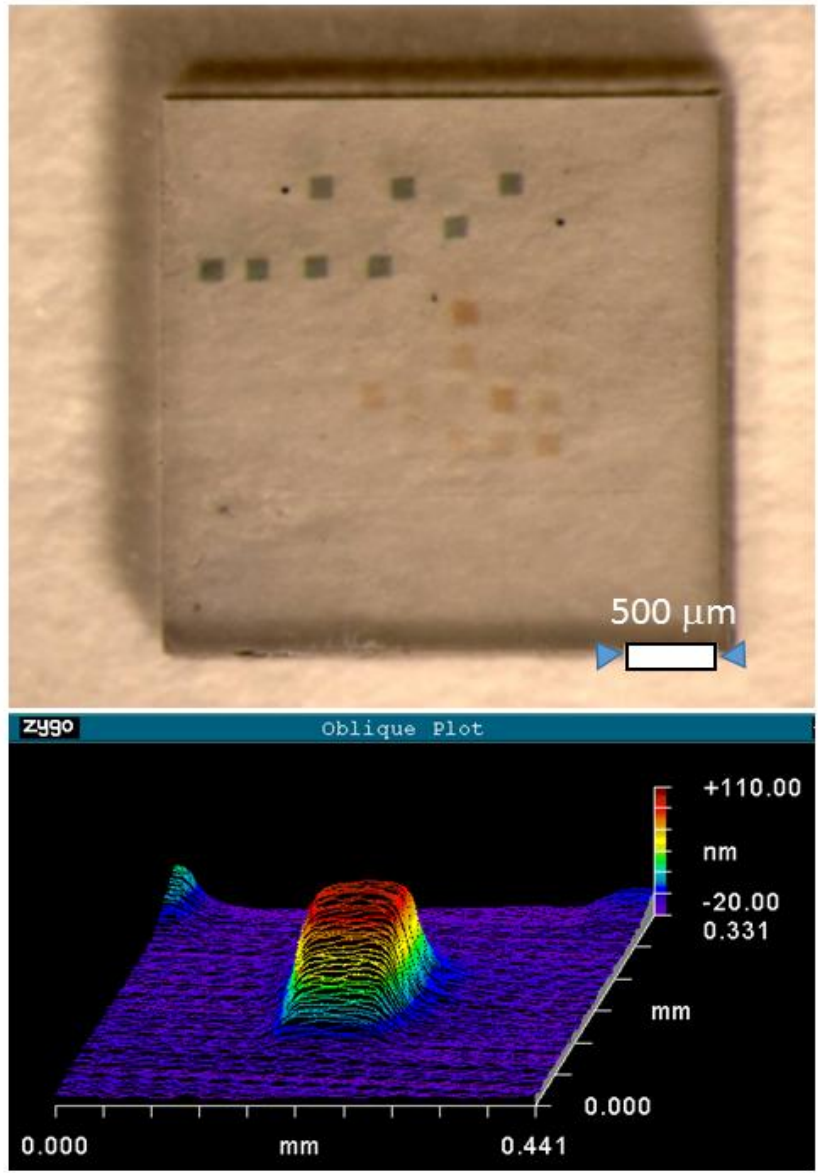

Fig. 4: a) *Optical micrograph of a CVD diamond sample irradiated with 2 MeV H ions. Increasing implantation fluence corresponds to increasing area opacity.* b) *Example of a measured swelling profile using white-light interferometry.*

## 4. Data analysis and discussion

All step height measurements for the different swift heavy ion implantations are shown in Fig. 5 as a function of the irradiation fluence. Data relevant to light ion implantations are also included for the sake of comparison. It is apparent that the same level of surface swelling, i.e. crystal amorphization, is achieved for considerably lower fluences in the case of swift heavy ions.

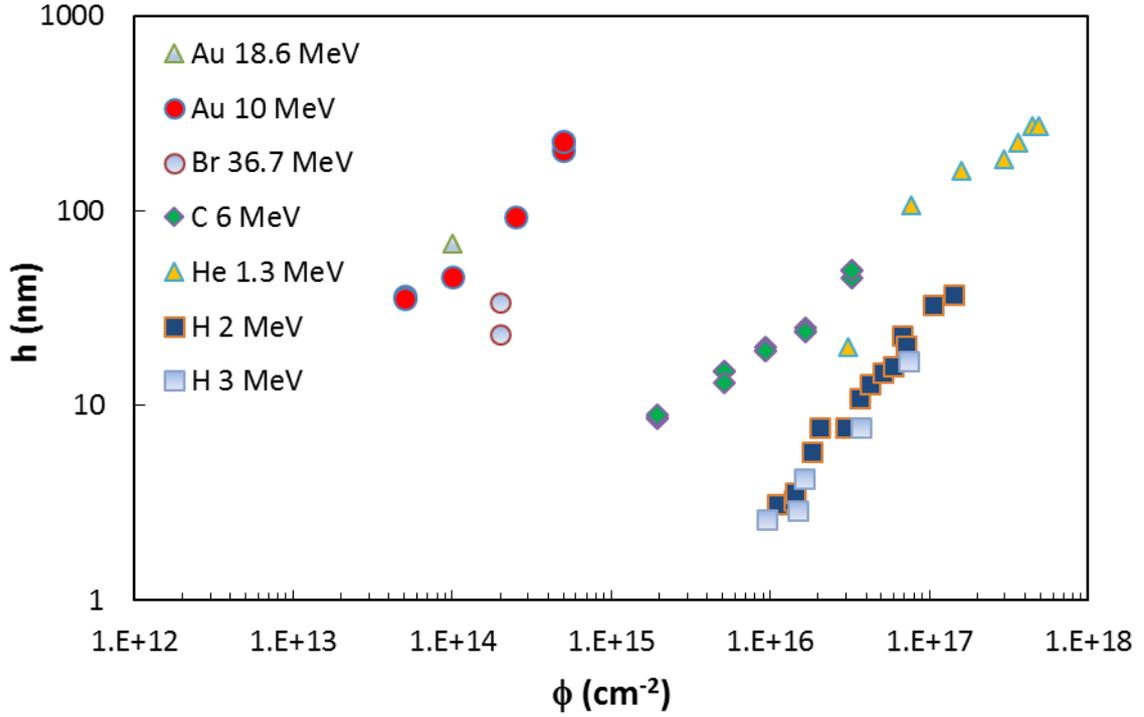

Fig. 5: *Swelling of irradiated areas of single-crystal diamond as a function of fluence Φ after irradiation with swift heavy ions and light ions.*

To compare the different implanted ion species and energies, the corresponding nuclear stopping power can be considered. A first estimation (which will be referred as "model 1" in the following) can be obtained by assuming a proportionality of the expected step height $h$ with the product between the irradiation fluence $\Phi$ and the maximum value of the nuclear stopping power along the ion trajectory $S_{n,max}$ according to SRIM [20, 21] simulations:

$$h = k \cdot \Phi \cdot S_{n,\max} \qquad (1)$$

The physical arguments behind this model are evidently oversimplified, the motivation being to explore the main trends with a very simple calculation tool, which can be used as a very rough estimate of the degree of damage produced in the sample. The "$h$ vs $\Phi\, S_{n,max}$" data for both swift heavy ions and light ions are plotted in Fig.6a, together with additional data for other ion species and energies taken from the literature, in similar implantation conditions (i.e. Room Temperature implantations) [23-26]. Best fits for data reported in Fig. 6a yield values of $k$ ranging from $2\times10^{-14}$ to $1.5\times10^{-12}$ nm$^2$cm$^2$ keV$^{-1}$, with a high scatter, so that this approximation is clearly unsatisfactory for fitting all of the data. A curve for $k = 3\times10^{-14}$ nm$^2$cm$^2$ keV is reported in Fig. 6a for reference.

A more accurate approach ("model 2") can be adopted that takes into account damage saturation effects occurring at high fluences [10]. According to this model, the structural damage at a given depth can be described as an exponential function of the product between fluence $\Phi$ and the nuclear stopping power $S_n(z)$. This model leads to an estimation of the damage fraction (i.e. the fraction of amorphized material) at a given depth $z$ in the sample as:

$$D(z) = 1 - \exp[-\Phi \cdot S_n(z)/a] \qquad (2)$$

with $a = (1.19 \pm 0.11)\times10^{15}$ cm$^{-2}$ keV nm$^{-1}$ determined in the case of swift heavy ion irradiations [10]. The total swelling at the surface of the sample $h$ can be then considered as proportional to the integral of the damaged fraction ($D$) over the full depth affected by the ion irradiation (model 2). Therefore, the step height can be expressed as:

$$h = b \int_0^{z\max} D(z)\,dz \qquad (3)$$

where the integral runs from the sample surface ($z = 0$) to the maximum ion range ($z = z_{max}$), and $b$ is a dimensionless constant, to be extracted from experimental results. The physical meaning of $b$ can be understood by observing that for the case in which the target is fully amorphized along the whole irradiation depth (i.e. $D(z) = 1$), the integral in Eq. (3) equals to $z_{max}$, so that $b$ can be interpreted as the linear expansion coefficient of the amorphized material in the direction normal to the surface. Fig. 6b shows the results of a linear fit of the "h vs $\int D(z)dz$" data, yielding a $b = 0.25\pm0.05$ estimation. Remarkably, the observed trend is satisfactorily consistent for different implantation conditions, including both light MeV ions and swift heavy ions.

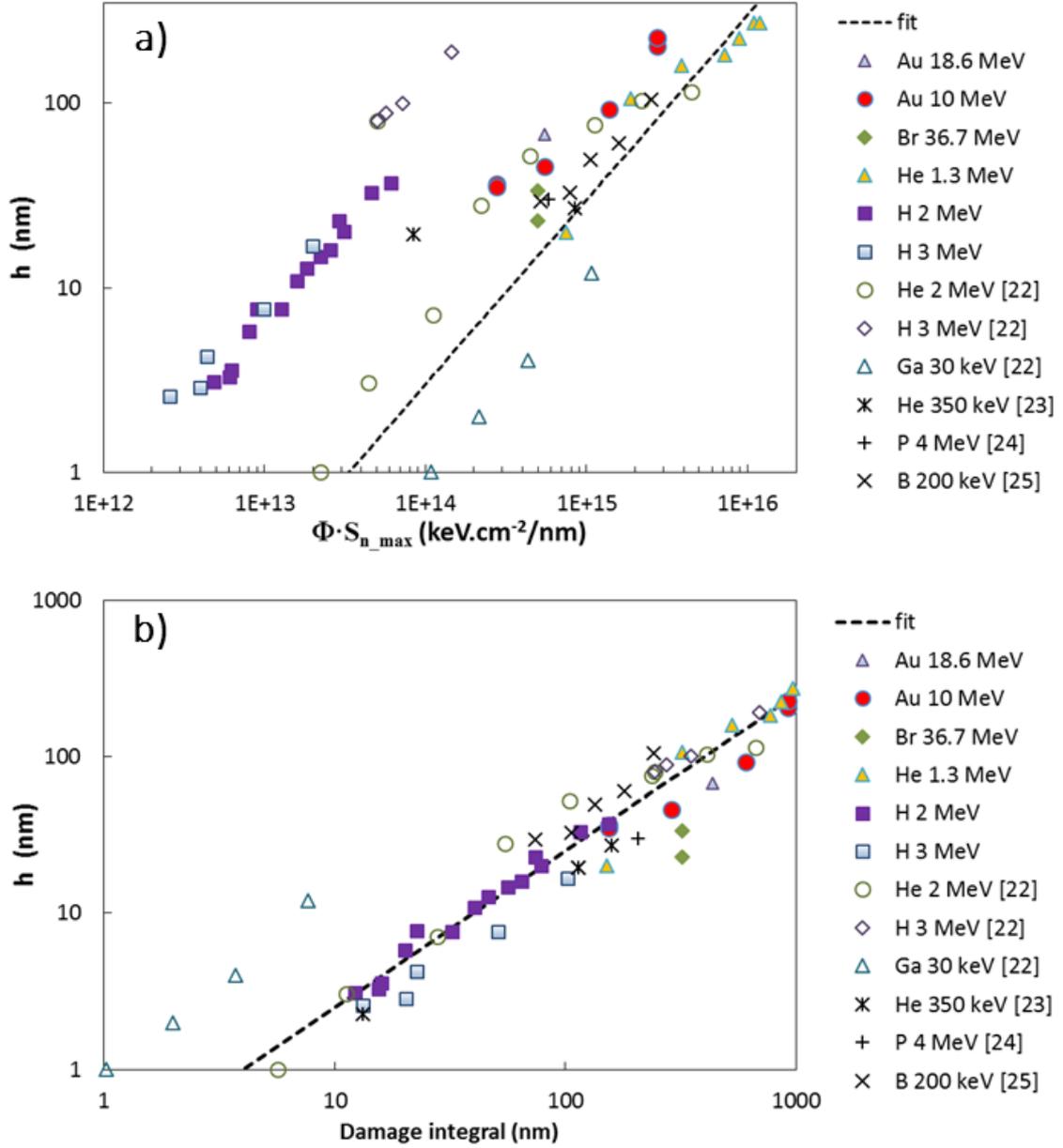

Fig. 6: *Swelling h of irradiated areas of single-crystal diamond as a function of either a) $\Phi \cdot S_{n\_max}$ or b) the damage fraction integral (see Eq. (3)) for irradiations with swift heavy ions and light MeV ions (additional data from the literature are included [23-26]). Fits on the data are included.*

The possible remaining cause for discrepancy from the predicted behaviour is related to the state of stress of the amorphized diamond region, which is not accounted for in the model. It is anticipated that for deeper implantations, the expanding amorphized region is subject to larger reaction stresses, and therefore deforms less, giving rise to smaller surface swelling. To capture this effect, and provide more reliable predictions of the expected mechanical deformation for a given implantation geometry, a numerical model based on Finite Element simulations is also adopted. The input structural/mechanical properties of damaged diamond are estimated using a phenomenological model based on vacancy density variation [22], instead of nuclear stopping power, as done in the previous section. This model also accounts for damage saturation effects at high fluences due to defect recombination in the crystal, so that the vacancy density $\rho_V$ in the depth direction $z$ can be expressed as:

$$\rho_V(z) = \alpha \cdot \left(1 - \exp\left[-\Phi \cdot \lambda(z)/\alpha\right]\right) \tag{4}$$

where $\lambda(z)$ is the linear vacancy density calculated numerically using the SRIM code, and $\alpha$ is the saturation vacancy density, estimated as $\alpha = 7.3 \times 10^{22}$ cm$^{-3}$ [27]. The corresponding mass density variation $\rho(z)$ can thus be written as:

$$\rho(z) = \rho_d - (\rho_d - \rho_{aC}) \cdot \left(1 - \exp\left[-\Phi \cdot \lambda(z)/\alpha\right]\right) \tag{5}$$

where $\rho_d = 3.52$ g cm$^{-3}$ is the density of diamond and $\rho_{aC} = 2.06$ g cm$^{-3}$ is the density of amorphous carbon [28]. The Young's modulus dependence on the vacancy density can be derived from Quantized Fracture Mechanics [29] for the case of single isolated (non-interacting) vacancies:

$$E(x) = E_d \cdot \left(1 - \kappa \frac{\rho(x)}{\rho_d}\right) \tag{6}$$

where $\kappa = 4.46$ is an empirical factor related to defect shape and interaction, derived from fitting Eq. (6) with results from ab initio simulations of diamond cells with varying vacancy densities [28].

Finite Element Model (FEM) simulations were carried out using the Structural Mechanics module of the COMSOL Multiphysics 5.0 package [30]. A 3-D model of the implanted diamond sample was created and the constrained expansion of the implanted diamond region due to the local density reduction was simulated. The latter was numerically modelled according to elasticity theory by introducing residual strains $\varepsilon_i(x)$ in the three principal directions of the implanted material ($i = 1, 2, 3$):

$$\varepsilon_i(z) = \sqrt[3]{\frac{\rho_d}{\rho(z)}} - 1 \qquad (7)$$

The local variations in material density and Young's modulus are determined using Eqs. (5) and (6). The same functional form, based on a rule of mixtures approach, is assumed for the variation of the Poisson's ratio:

$$v(z) = v_d - (v_d - v_{aC})(1 - \exp[-\Phi \cdot \lambda(z)/\alpha]) \qquad (8)$$

where $v_d = 0.07$ and $v_{aC} = 0.34$ are the Poisson's ratios of pristine diamond and amorphous carbon, respectively [31]. Figure 7 shows results of a typical 3D FEM simulation for the swelling of an Au 10 MeV implanted area. The swelling pattern correctly predicts the experimentally observed ~2 μm width of the slope between irradiated and unirradiated areas.

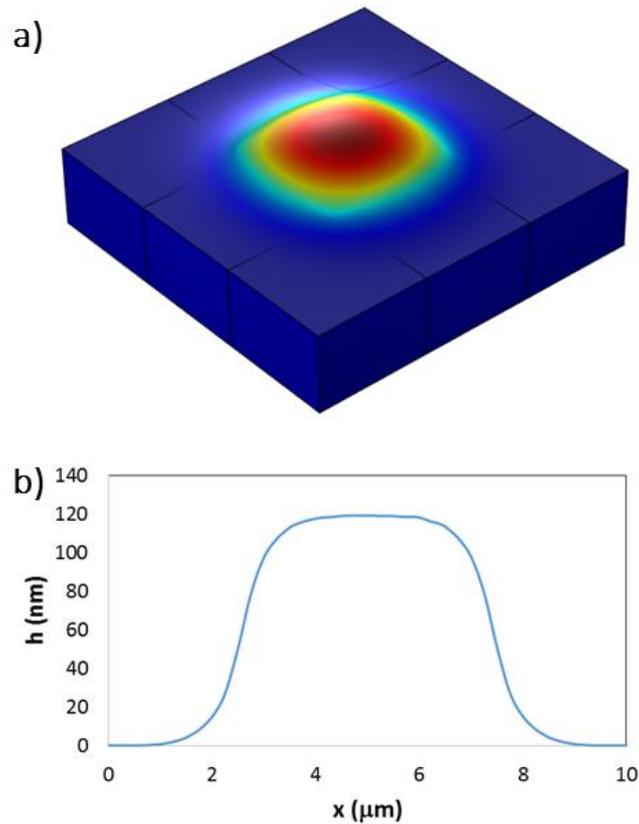

Fig. 7: *a) Simulated FEM surface swelling for a single square area implanted with 10 MeV Au ions (F = 2.5×10$^{14}$ cm$^{-2}$). b) Corresponding calculated swelling line profile*

The predictions of FEM simulations are compared in Figure 8 with those of models 1 and 2 and with measured data (using profilometry and AFM) for one of the swift heavy ion implantations (i.e. 10 MeV Au). As expected, the FEM simulations provide the best fit on the experimental data (with a standard deviation of $\sigma_d = 18$ nm) with respect to model 1 ($\sigma_d = 88$ nm) and 2 ($\sigma_d = 37$ nm), while it is worth remarking that they do not require the introduction of fitting parameters such as $k$ or $b$. Actually, FEM simulations could indeed be used to determine the above-mentioned $k$ or $b$ values, i.e. to calibrate the models in the absence of experimental data. In addition, FEM calculations allow the determination of full-field deformation profiles and internal stresses in the material, which can be useful, e.g. to predict microcracking effects.

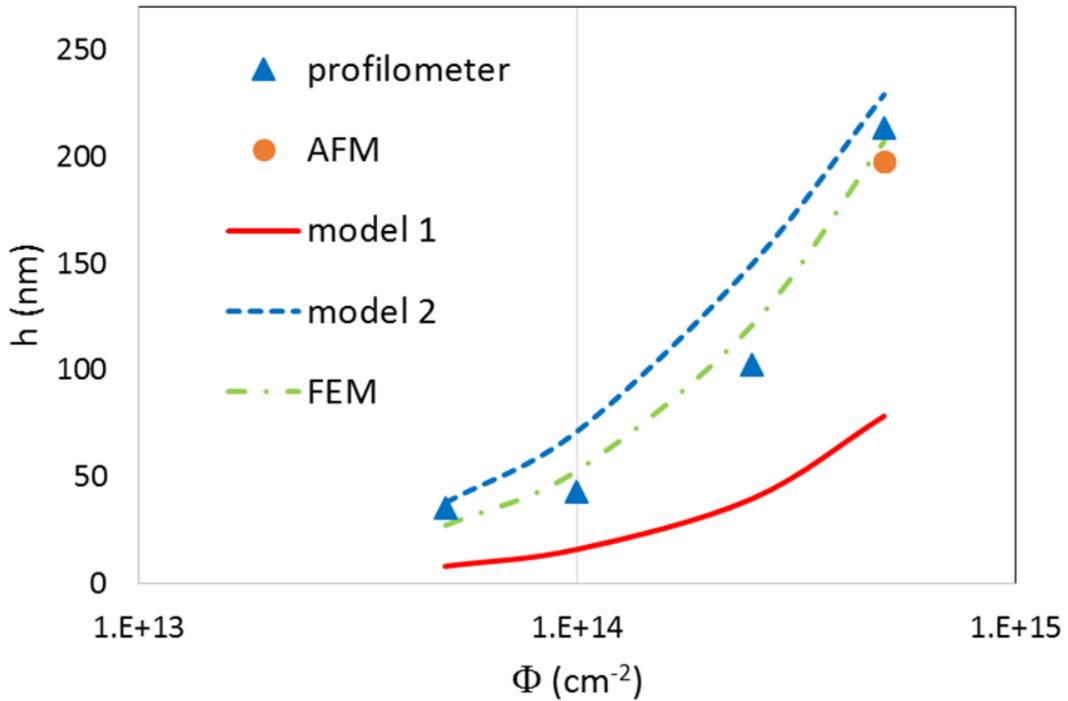

Fig. 8: *Step height h between irradiated and unirradiated areas of single-crystal diamond, after masked irradiation with a 10 MeV Au ion beam. Experimental measurements are derived from profilometry and AFM. Simulations are performed using the two proposed phenomenological models ("model 1" and "model2"), as well as FEM analysis ("FEM"). The values of the free parameters are $k = 3 \times 10^{-14}$ nm$^2$ keV$^1$ cm$^{-2}$ (model 1) and $b = 0.25$ (model 2).*

5. **Conclusions**

Surface swelling effects induced in single-crystal diamond by buried structural damage generated by ion irradiation have been analysed and discussed, presenting for the first time data relative to swift heavy ions. It is shown that the behaviour already observed for light MeV ions also occurs for swift heavy ion irradiation. In this case, severe swelling effects develop for considerably lower ion fluences, in the range $10^{13}$ to $10^{15}$ cm$^{-2}$. Data relative to heavy and light ions were analysed and described using three alternative models. The first model is based on the maximum stopping power and fluence for the given implantation (Eq. 1). It provides a very simplified phenomenological view of the damage process and cannot describe the data accurately, but can still be useful as an easily-calculated first guess of the predicted swelling one can expect for given irradiation conditions. The second model is based on the damage depth profile as given by SRIM and captures most of the details, including damage accumulation and saturation, while still being relatively easy to calculate (see Eqs. 2 and 3). Finally, a more accurate description has been obtained by FEM analysis, allowing the description of the mechanical deformation due to the specific boundary conditions of the geometry of the irradiated areas.

In addition, the measurements presented in this paper illustrate how the swelling effect can be used to generate customized surface landscapes, formed by lightly damaged single-crystal material, due to the force exerted by the heavily damaged volumes buried several microns beneath. With proper irradiation parameters and masks these landscapes can be patterned in a customized way. The models presented in this paper represent a useful tool to determine the correct irradiation parameters necessary to design predefined patterns on a diamond single-crystal surface with specific functional properties, such as controlled wettability, increased cell adhesion, etc.

**Acknowledgements**

GG acknowledges support from the ALBA synchrotron, W. Schildkamp for inspiring discussions on the behavior of diamond and J. Ferrer for his help in experiment preparation.
GG, MD-H, VT-M, OP-R and JO acknowledge the projects MAT-2011-28379-C03-02 of the Spanish Ministry of Economy and Competitiveness, TECHNOFUSION(II)CM (S2013/MAE2745) of the Community of Madrid, and Moncloa Campus of International

Excellence (UCM-UPM) foundation for offering a PICATA postdoctoral fellowship (OP-R).

FP is supported by the "DiNaMo" project n° 157660 funded by National Institute of Nuclear Physics. PO is supported by the FIRB "Futuro in Ricerca 2010" project (CUP code: D11J11000450001) funded by MIUR and by the "A.Di.N-Tech." project (CUP code: D15E13000130003) funded by the University of Torino and "Compagnia di San Paolo". The MeV ion beam implantations performed at the INFN Legnaro National Laboratories was supported by the "Dia.Fab." experiment, and those at the INFN LABEC Laboratory by the "FARE" and "CICAS" experiments.

NMP is supported by the European Research Council (ERC StG Ideas 2011 BIHSNAM n. 279985, ERC PoC 2013-2 KNOTOUGH n. 632277 and ERC PoC 2015 SILKENE no. 693670), by the European Commission under the Graphene Flagship ("Nanocomposites", n. 604391). FB acknowledges support from BIHSNAM.

LL-M and CO acknowledge the Spanish MINECO through the Severo Ochoa Program (SEV-2015-0496) and MAT2013-47869-C4-1-P.

CO acknowledges the specific agreement between ICMAB-CSIC and the Synchrotron Light Facility ALBA.